\documentclass[final,sort&compress ]{aipproc}

\layoutstyle{6x9}

\begin{document}

\title{Determining orbits for the Milky Way's dwarfs}


\author{H. Lux}{
  address={Department of Theoretical Physics, University of Z\"urich, Winterthurerstr. 190, CH-8057 Z\"urich, Switzerland}
  ,altaddress={lux@physik.uzh.ch} 
}

\author{J. I. Read}{
  address={Department of Physics \& Astronomy, University of Leicester, University Road, Leicester,  LE1 7RH, United Kingdom.}
}

\author{G. Lake}{
  address={Department of Theoretical Physics, University of Z\"urich, Winterthurerstr. 190, CH-8057 Z\"urich, Switzerland}
}

\begin{abstract}
We calculate orbits for the Milky Way dwarf galaxies with proper motions, and compare these to subhalo orbits in a high resolution cosmological simulation. We use this same simulation to assess how well are able to recover orbits in the face of measurement errors, a time varying triaxial gravitational potential, and satellite-satellite interactions. We find that, for present measurement uncertainties, we are able to recover the apocentre $r_a$ and pericentre $r_p$ to $\sim 40$\%.  However, even with better data the non-sphericity of the potential and satellite interactions during group infall make the orbital recovery more challenging. Dynamical friction, satellite mass loss and the mass evolution of the main halo play a more minor role. 

We apply our technique to nine Milky Way dwarfs with observed proper motions. We show that their mean apocentre is consistent with the most massive subhalos that form before $z=10$, lending support to the idea that the Milky Way dwarfs formed before reionisation. 
\end{abstract}

\maketitle


\paragraph{Testing the method}
In \citep{MYPAPER} we derive orbits for the Milky Way's (MW) dwarfs with proper motion data, by integrating their orbits backwards in a fixed potential. We test this procedure using the high resolution Via Lactea I cosmological simulation of a MW analogue \citep[VL1; ][]{2007ApJ...657..262D}.  For this we extract from the simulation two sets of subhalos: the 50 most massive today ($z^{50}_0$), and the 50 most massive before redshift $z=10$ ($z^{50}_{10}$)\footnote{See the data at http://www.ucolick.org/~diemand/vl/ }. In both cases, we include only subhalos with mass $M > 10^7 M_\odot$ and distance to the centre of the main halo $r< 150$\,kpc at redshift $z=0$.

To test the limitations of this approach, we integrate the orbits in 3 different models:  {\bf The fiducial model}: This uses a static, spherical, NFW potential \citep{1996ApJ...462..563N} fit to the VL1 main halo at $z=0$ \citep{2007ApJ...657..262D}; {\bf The dynamical friction model }: This uses the same potential with dynamical friction forces \citep{1943ApJ....97..255C}, a prescription for the mass loss of satellites as well as the mass growth of the main halo \citep{2004MNRAS.351..891Z};  {\bf The triaxial model }: This uses a triaxial NFW potential as in \citep{2007ApJ...671.1135K,2009ApJ...702..890G}.  

We find that the orbits without measurement errors are systematically effected by an incorrect halo shape as well as satellite-satellite interactions. Dynamical friction, mass loss of the satellite and mass evolution of the main halo play a minor role. Current measurement errors are more significant than model systematics. They bias the mean apo-/pericentre to higher values. We are able to recover apo-/pericentre up to 40\%. 

\begin{figure*}
\centering
\includegraphics[width=0.33\textwidth]{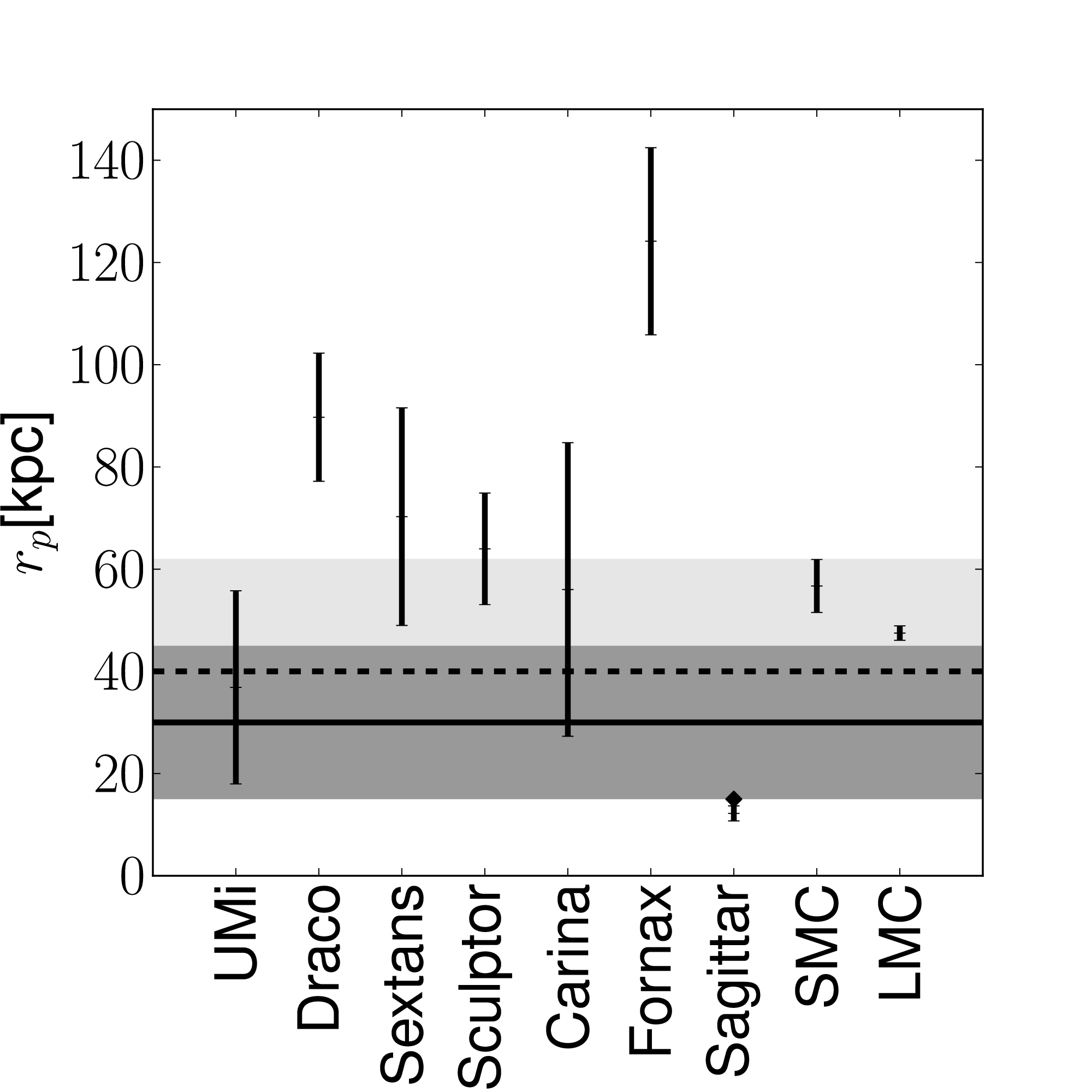}
\includegraphics[width=0.33\textwidth]{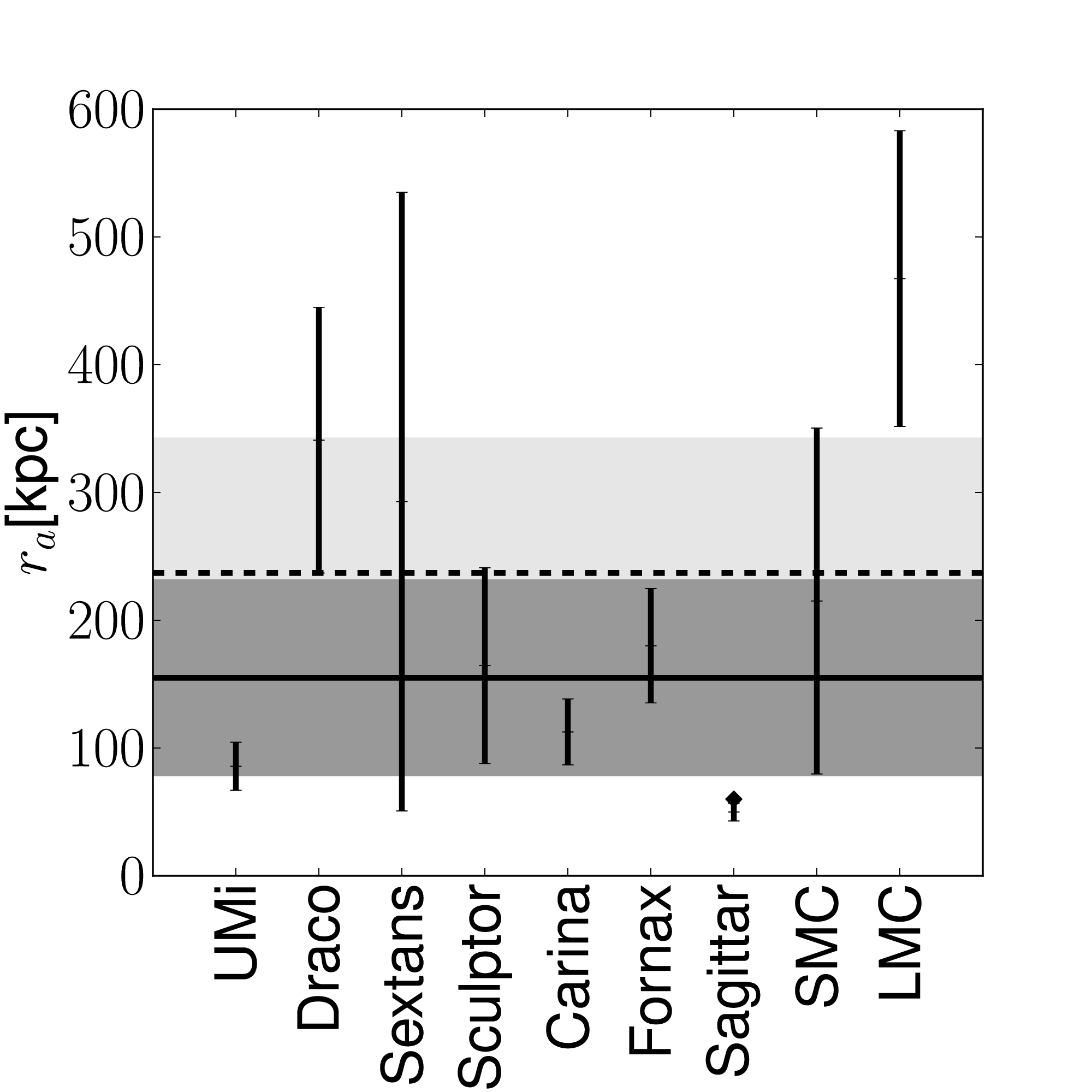}
\includegraphics[width=0.33\textwidth]{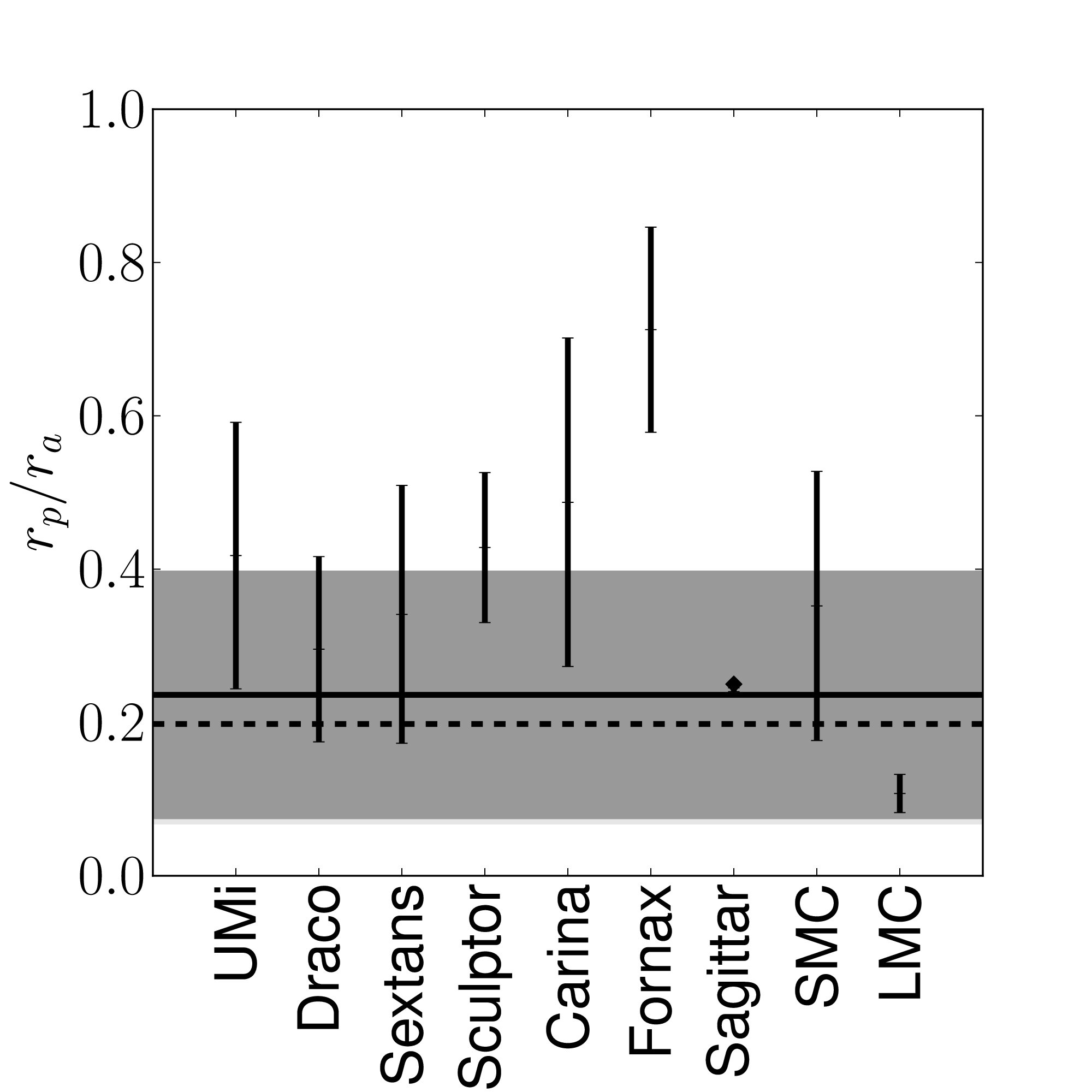}
\caption{Recovered peri-/apocentres for nine Milky Way dwarfs with observed proper motions. The error bars show the values for the oblate potential of \cite{2005ApJ...619..807L}. Overlaid are the mean and standard deviation of the values from $z^{50}_0$ (light grey band, dashed line) and $z^{50}_{10}$ (dark grey band, solid line). The black diamonds denote the values derived from the Sagittarius stream \citep{2005ApJ...619..807L}.}
\label{fig:dwarfs}
\end{figure*}

\paragraph{Milky Way Dwarfs} We derive orbits for the MW dwarfs with proper motion data, by integrating their motion backwards in the oblate potential from \cite{2005ApJ...619..807L}. We take current measurement errors into account by building an ensemble of 1000 orbits for each dwarf drawn from its error distributions. The results are shown in Figure \ref{fig:dwarfs}. Our recovered mean pericentre $\langle r_p \rangle$ is higher than the mean pericentre in the simulation. This could be either due to satellite depletion by the galactic disk as recently discussed by \citep{2009arXiv0907.3482D}; but can also be explained by the bias we see in the VL1 recovered data which is due the large proper motion errors. This can be reduced by better quality data. The apocentre distances $r_a$ instead are lower than the mean of the $z^{50}_0$ sample, but consistent with the $z^{50}_{10}$ sample. This lends further support to the idea that the MW's dwarfs formed early before reionisation \cite[e.g.][]{2004ApJ...609..482K,2006MNRAS.368..563M,2009arXiv0903.4681M}.

\bibliographystyle{aipproc}   

\bibliography{001hlux}

\end{document}